\documentclass[12pt]{article}
\usepackage{graphicx}
\usepackage{latexsym}

\newcommand{\EQ}{\begin{equation}}
\newcommand{\EN}{\end{equation}}
\newcommand{\bea}{\begin{eqnarray}} 
\newcommand{\ena}{\end{eqnarray}}
\newcommand{\vs}[1]{\vspace{#1 mm}}

\renewcommand{\d}{\delta}
\newcommand{\e}{\epsilon}

\def\bbox{{\,\lower0.9pt\vbox{\hrule \hbox{\vrule height 0.2 cm
\hskip 0.2 cm \vrule height 0.2 cm}\hrule}\,}}
\newcommand{\dsl}{\pa \kern-0.5em /}

\newcommand{\pa}{\partial}

\renewcommand{\k}{\kappa}

\newcommand{\nn}{\nonumber\\}
\newcommand{\p}[1]{(\ref{#1})}

\begin{document}

\topmargin 0pt
\oddsidemargin 0mm
\renewcommand{\thefootnote}{\fnsymbol{footnote}}
\begin{titlepage}

\setcounter{page}{0}

\vs{10}
\begin{center}

{\Large\bf Page charge of D-branes and its
behavior in topologically nontrivial B-fields}

\vs{15}

{\large 
Jian-Ge Zhou\footnote{e-mail address: jgzhou@hep.itp.tuwien.ac.at}}

\vs{10}
{\em Institut f\"ur Theoretische Physik, Technische Unversit\"at Wien,\\
Wiedner Hauptstrasse 8-10/136, A-1040 Wien, Austria} \\
\end{center}

\vs{15}
\centerline{{\bf{Abstract}}}
\vs{5}
The RR Page charges for the D(2p+1)-branes with B-field in
type IIB supergravity are constructed consistently from
brane source currents. The resulting Page charges are
B-independent in nontrivial and intricate way. It is
found that in topologically trivial B-field the Page
charge is conserved, but in the topologically nontrivial
B-field it is no longer to be conserved, instead there is a 
jump between two Page charges defined
in each patch, and we interpret this jump as Hanany-Witten effect.

\end{titlepage}
\newpage

\renewcommand{\thefootnote}{\arabic{footnote}}
\setcounter{footnote}{0}

{\Large\bf 1. Introduction}
\vs{5}

Supergravity theories contain two types of Chern-Simons terms:
one involves the wedge product of one potential with any
number of field strengths, and the other appears in the
kinetic term for the modified field strengths, which result in that
the equations of motion for gauge fields are nonlinear in the
gauge fields and the standard Bianchi identities turn into modified
ones~\cite{jp}. These peculiar features make the definition of charge
for gauge fields in supergravity theories much more subtle, 
actually, this is central issue raised by 
Bachas, Douglas and Schweigert in~\cite{bds} where they
found that the RR charges of the D2-branes in group manifold 
are irrational due to the integral $\int B$.
In~\cite{wt}, it was argued that the D0-brane
charge arising from the integral over the D2-brane of the
pullback of the B field is cancelled by the bulk 
contributions\footnote{In~\cite{amm},
it rephrased the reasoning of~\cite{wt} and showed
that if one writes the Wess-Zumino action in
terms of the gauge invariant field strength, the resulting
RR brane charges are independent of B fields, but in their
construction it is not clear what properties these RR charges
possess.}, so only $\int F$ should be quantized.  In~\cite{dm}, 
three notions of charge for D-brane in type IIA supergravity 
were proposed, and of particular interest are brane source charge
and Page charge, the first
is gauge invariant and localized, but not conserved,
the second one is conserved, localized and invariant under
small gauge transformation\footnote{Since brane source charge
and Page charge were defined separately in~\cite{dm},
it is unclear how to relate one to other which is essencial 
to the question whether it is possible to consistently 
define the general RR Page charges for a D2p-brane from
brane source charges.}. In~\cite{jgz}, the B-independent RR charges
of D2p-branes with B fields in type IIA supergravity were constructed
explicitly from brane source charges by exploiting
the equations of motion and the nonvanishing 
(modified) Bianchi identities, and were identified with Page charge
from their properties  --- conserved and localized. 

In type IIB supergravity there is no covariant action due to
the self-dual field strength $\tilde{F}_{5} = *\tilde{F}_{5}$,
also because of the $SL(2,R)$ symmetry of the type IIB theory,
the $\tilde{F}_{5}$ is defined as $F_{5} + \frac{1}{2}C_{2}\wedge H
- \frac{1}{2}B\wedge F_{3}$, but in type IIA supergravity the
physical gauge invariant field strength is rather defined as 
$\tilde{F}_{2p+2} = F_{2p+2} - C_{2p-1}\wedge H$. 
When 11D $N=1$ supergravity is dimensionally reduced by
compactification on a small circle in the $X^{11}$ direction,
the resulting theory is type IIA supergravity, and
11-momentum turns to the D0-brane charge, which is much
simpler than the relation between 11D $N=1$ supergravity 
and type IIB supergravity.
Due to these special features in type IIB supergravity, 
the approaches in~\cite{wt},\cite{amm},\cite{dm} can not be
applied to type IIB case\footnote{For instance,
Marolf's argument that Page charges are independent of B-field~\cite{dm}
does not hold in type IIB case.}, so it would
be interesting to check whether it is possible 
to consistently construct Page charges for D(2p+1)-branes in 
B fields along the line of~\cite{jgz} and whether the 
resulting RR Page charges are B-independent or not. 
In~\cite{jgz}, it
was shown that the conserved Page charge can be defined consistently
only if $H = dB$ holds in the whole region. When it does not
hold, for instance, in the background of NS5-branes, the NS B-field
is topologically nontrivial and we have to cover $S^{3}$ with two
patches, in each patch we can construct conserved Page charge, then
one may ask how these two RR Page charges are connected and
how to interpret this physical phenomenon.
Because the simple example which encodes brane
creation phenomenon~\cite{hw}-~\cite{kz} is the D3-brane probe in
the background of $\kappa$ coinciding NS5-branes~\cite{op},
the further motivation of studying RR Page charges
in type IIB supergravity is to survey which charge nonconservation
is responsible for the Hanany-Witten effect. 

Motivated by the above, in the present paper 
we discuss in type IIB supergravity how to construct
B-independent RR charges for D(2p+1)-branes in
B-field from brane source charges. Starting from
IIB supergravity plus D-brane sources, we derive the equations
of motion and the nonvanishing (modified) Bianchi identities that
define the duals of brane source currents for D(2p+1)-branes.
We insert the equations for the
duals of brane source currents for D(2p+1+2n)-branes into
that for the D(2p+1)-brane iteratively, and find that the resulting
equations can be recast into the form whose left sides of equations
are exterior derivative and the right sides are localized objects
which indicates that the right side localized objects can be identified
as Page charges because of their conservation and locality. Plugging
the brane source charges into the expression of the Page charges,
we find that all the Page charges are independent of the background
B fields, actually it is highly nontrivial and intricate
that all the B-dependent terms from different sources
are exactly cancelled with each other. However in the construction
of the Page charges, we assume that in the whole region the relation
$H = dB$ holds, and resulting Page charges are conserved in
this region. To see how Page charges behave in topologically
nontrivial B-field, we consider D3-brane probe in the
background of $\kappa$ coincident NS5-branes where in the spherical
coordinates we have $dH = 0$, but $\int\limits_{S^3} H \neq 0$.
With this particular example, we show that the nonconservation of brane
source charge does not describe the Hanany-Witten effect, it
only reflects the fact that the brane source charge depends on
the background nonconstant B-field, thus we clear up the puzzle
raised in~\cite{dm}.  
In the topological nontrivial B-field, we show that the Page charge 
is no longer
conserved, instead there is a jump between two Page charges defined
in each patch. We interpret this jump as Hanany-Witten effect,
in other words, we find a new way to describe brane creation
phenomenon.
 
The paper is organized as follows. In section 2, 
the Page charges for D(2p+1)-branes in topologically trivial
 B-field are constructed
explicitly from brane source charges by exploiting
the equations of motion and the nonvanishing 
(modified) Bianchi identities. It is shown that all
Page charges are independent of B fields. In section 3, 
the D3-brane probe in the background of $\kappa$ coincident NS5-branes 
is discussed, the Page charge of D1-branes is calculated.
We find there is a jump between the Page charges constructed
in each patch, and interpret it as Hanany-Witten effect.
The nonconservation of brane source charge is attributed
to the fact $dB \neq 0$.
In section 4, we present our summary and conclusion.

\vs{5}
{\Large\bf 2. Page charges for D(2p+1)-branes in B fields}
\vs{2}

For IIB supergravity in ten dimensions there is no covariant
action due to the self-dual field strength 
$\tilde{F}_{5} = *\tilde{F}_{5}$, but the field equations from the
following action are consistent with 
$\tilde{F}_{5} = *\tilde{F}_{5}$~\cite{jp}
\bea
S_{IIB}&=&\frac{1}{2\k_{10}^{2}}\int d^{10}x\sqrt{-G} 
\Big\{e^{-2\Phi}(R + 4\pa_{\mu}\Phi\pa^{\mu}\Phi - \frac{1}{2}|H|^{2})\nn
&&
-\frac{1}{2}(|F_{1}|^{2} + |\tilde {F}_{3}|^{2}
+ \frac{1}{2}|\tilde {F}_{5}|^{2})\Big\}
-\frac{1}{4\k_{10}^{2}}\int C_{4}\wedge H\wedge F_{3}
\ena
where $\k_{10}^{2}$ is the gravitational coupling in
ten dimensions, and the field strengths $H, F_{1}, F_{3}, F_{5}$,
$\tilde {F}_{3}$, and $\tilde {F}_{5}$ are defined as\footnote{There is
minus sign difference for the field B between our notation and
that in~\cite{jp}.} 
\EQ
H = dB,~F_{1} = dC_{0},~F_{3} = dC_{2},~F_{5} = dC_{4},~ 
\tilde {F}_{3} = dC_{2} - C_{0}\wedge H,~ 
\tilde {F}_{5} = dC_{4} + \frac{1}{2}C_{2}\wedge H
- \frac{1}{2}B\wedge F_{3}
\EN
The field equations from the above action are consistent with 
$\tilde{F}_{5} = *\tilde{F}_{5}$, but they do not imply it, this
must be imposed as an added constraint on the solutions, and
it cannot be imposed on the action or else the incorrect equations
of motion result. This formulation is satisfactory for a classical
treatment, however it is not simple to impose the contraint in the 
quantum theory but this will not be important for our following
purposes. The above action can be recast into
\bea
S_{IIA}&=&\int d^{10}x\sqrt{-G} 
e^{-2\Phi}(R + 4\pa_{\mu}\Phi\pa^{\mu}\Phi)\nn
&&
- \frac{1}{2}\int \Big\{-e^{-2\Phi} H\wedge *H + F_{1}\wedge *F_{1}\nn
&&
+ \tilde {F}_{3}\wedge *\tilde {F}_{3} + \frac{1}{2}\tilde {F}_{5}
\wedge *\tilde {F}_{5})
+ C_{4}\wedge H\wedge F_{3}\Big\}
\label{2a}
\ena
where we have chosen $2\k_{10}^{2} = 1$ so that
the kinetic term is canonical.

A D(2p+1)-brane in the type IIB supergravity background has
a world-volume action given by a sum of Born-Infeld and
Wess-Zumino terms
\EQ
S_{D-brane} = S_{BI} + S_{WZ}
\EN
The Born-Infeld part of the action is 
\EQ
S_{BI} = -T_{2p+1}\int e^{-\Phi}\sqrt{-det(G_{ab} + B_{ab} + F_{ab})}
\EN
where $G_{ab}$, $\Phi$, $B_{ab}$ are the pullback of 
spacetime metric, dilaton and Neveu-Schwarz two form B to the D(2p+1)-brane
world-volume, $F_{ab}$ is the field strength of U(1) gauge field
living on the brane.
The Wess-Zumino (WZ) terms couple the brane to the spacetime RR gauge
field through
\EQ
S_{WZ} = \int ( \sum\limits_{i} C^{(i)})\wedge e^{B+F}
\EN

\vs{5}
{\Large\bf 2.1 D1-brane in B fields}
\vs{2}

In this subsection we discuss D1-brane in B fields, the 
Wess-Zumino term for D1-brane is
\EQ
S_{WZ}^{D1}= \int\Big\{C_{2} + C_{0}\wedge (B+F)\Big\}
\EN
which indicates that there are D-instantons living on the D-string.
Since D-branes are sources of RR gauge field, we consider
IIB supergravity + the coupled D-brane source, and see how 
D-string, D-instantons induce $C_{2}$, $C_{0}$ gauge field,
which can be described by the equations of motion.
As the duals of brane source currents arises from varying the brane action
with respect to the gauge field, from the total action 
$S_{T} = S_{IIB} + S_{D1}$ we get the equations of
motion for $C_{0}$ and $C_{2}$
\bea
d*F_{1} - H\wedge *\tilde{F}_{3} &=& -*j^{bs}_{D(-1)}\nn
d*\tilde{F}_{3}- H\wedge\tilde{F}_{5} &=&-*j^{bs}_{D1}
\label{eqm}
\ena
where in the derivation of the second equation of \p{eqm},
we have used the fact $H\wedge H = 0$. Since the brane
source current is gauge invariant the left side of the 
the second equation should be written in terms of the
physical field strengthes $\tilde{F}_{3}$ and $\tilde{F}_{5}$.
Generally speaking, we should write each equation of motion
for the gauge fields as a polynomial in the gauge invariant
improved field strengths, their hodge duals, and exterior
derivatives of these and then let the right hand side be some
$*j$ which are associated with some D-branes, NS5-branes
or fundamental strings. The brane source currents for D1-brane
and D-instantons are
\EQ
(j_{D1}^{bs})^{\mu\nu}(x) = \int dX^{\mu}(\xi)\wedge dX^{\nu}(\xi)
d^{10}(x-X(\xi))\nn
\EN
\EQ
(j_{D(-1)}^{bs})(x) = \int d^{2}\xi \e^{ab}
[B_{\mu\nu}(X(\xi))\pa_{a}X^{\mu}(\xi)\pa_{b}X^{\nu}(\xi) + F_{ab}(\xi)]
\d^{10}(x-X(\xi))
\EN
where $x^{\mu}$ are ten-dimensional coordinates, $X^{\nu}(\xi)$ are 
the embedding coordinates of D-string in ten dimensions, and
$\xi^{a}$ are two-dimensional brane worldvolume coordinates.
The brane source charge for D-instantons is 
\EQ
Q_{D(-1)}^{bs} = \int *j_{D(-1)}^{bs} = \int\limits_{V_2} (B + F)
\EN
where ${V_2}$ is two-dimensional Euclidean worldsheet of D-string.

For the convenience of the following discussion, we write down the
equation of motion (or modified Bianchi identity due to self-duality
of $\tilde{F}_{5}$) for $C_{4}$ and the (modified) Bianchi identity
for $\tilde{F}_{3}$ and $F_{1}$ without the brane sources for 
D3-, D5-, D7-branes
\EQ
d*\tilde{F}_{5} + H\wedge \tilde{F}_{3} = d\tilde{F}_{5} + 
H\wedge \tilde{F}_{3} = 0
\label{bi3}
\EN
\EQ
d\tilde{F}_{3} + H\wedge F_{1} = 0
\label{bi5}
\EN
\EQ
dF_{1} = 0
\label{bi7}
\EN
where we have impose the self-dual condition 
$*\tilde{F}_{5} = \tilde{F}_{5}$. If we rewrite IIB supergravity
in terms of $C_{6}$ and $C_{8}$ due to $C_{2}$ and $C_{0}$,
the Bianchi identities \p{bi5} and \p{bi7} turn into the equation
of motion for $C_{6}$ and $C_{8}$.
Since there are only D-string and D-instantons on the background
of B-field without any other brane sources, 
so we have the equations \p{bi3}--\p{bi7}. It is easy to
see $d*j_{D1}^{bs} = 0$, $d*j_{D(-1)}^{bs} \not= 0$, that is,
$Q_{D1}^{bs}$ is conserved but $Q_{D(-1)}^{bs}$ nonconserved.

Since the brane source charge $Q_{D(-1)}^{bs}$ is not conserved,
we explore whether it is possible to find some conserved
charge for D-instantons from brane source currents defined by
\p{eqm}, and \p{bi3}-\p{bi7}. In doing so, we first rewrite
the second term in the first equation of \p{eqm} $H\wedge*\tilde{F}_{3}$
as $d(B\wedge *\tilde{F}_{3}) - B\wedge d*\tilde{F}_{3}$,
then insert the second equation of \p{eqm} into the term 
$B\wedge d*\tilde{F}_{3}$, besides the exterior derivative
and localized term (in the following description, we do not
mention such terms), we have $B\wedge H\wedge *\tilde{F}_{5}$
which can be written as $\frac{1}{2}d(B\wedge B\wedge \tilde{F}_{5})
-\frac{1}{2}B\wedge B\wedge d\tilde{F}_{5}$. Plugging \p{bi3}
into $\frac{1}{2}B\wedge B\wedge d\tilde{F}_{5}$, we leave with
$\frac{1}{2}B\wedge B\wedge H\wedge \tilde{F}_{3}$ which is
recast into $\frac{1}{6}d(B\wedge B\wedge B\wedge \tilde{F}_{3})
- \frac{1}{6}B\wedge B\wedge B\wedge d\tilde{F}_{3}$.
We further insert \p{bi5} into 
$\frac{1}{6}B\wedge B\wedge B\wedge d\tilde{F}_{3}$ and
have the term $\frac{1}{24}B\wedge B\wedge B\wedge d\tilde{F}_{1}$.
Exploiting \p{bi7} and shifting the exterior derivative terms
on the left side of the equation and the localized terms on
the right side\footnote{This is our strategy to derive
RR Page charge for D-brane, in the following subsections we will 
exploit it.}, then we arrive at
\EQ
d\Delta_{D(-1)} = *j_{D(-1)}^{bs} - B\wedge *j_{D1}^{bs}
\label{dieq}
\EN
with
\EQ
\Delta_{D(-1)} = -*F_{1} + B\wedge *\tilde{F}_{3} - 
\frac{1}{2}B\wedge B\wedge\tilde{F}_{5} - 
\frac{1}{6}B\wedge B\wedge B\wedge\tilde{F}_{3}
- \frac{1}{24}B\wedge B\wedge B\wedge B\wedge F_{1}
\label{deltai}
\EN
In the above derivation, we have assumed that
the D1-brane stays in the region where the relation $H = dB$ holds.
Since Page charge is conserved and localized~\cite{dm},~\cite{jgz}, 
Eq. \p{dieq}
shows that the dual of the Page current for D-instantons are
\EQ
*j_{D(-1)}^{Page} = *j_{D(-1)}^{bs} - B\wedge *j_{D1}^{bs}
\label{dic}
\EN
Here we should mention that in~\cite{dm}, the brane source charge 
and Page charge were defined separately, the Page charge was given 
just by definition, they did not discuss how to define
Page charge based on the definition for brane source charge, but here 
\p{dic} is derived from the brane source currents.
The RR Page charge is 
\EQ
Q_{D(-1)}^{Page} = \int *j_{D(-1)}^{Page} = \int\limits_{V_{2}}F
\EN
which is independent of the background NS B-field. 
If we do T-duality along other extra four dimensions,
the D(-1)/D1 bound state turns into D3/D5 one, and
Eq.\p{dic} becomes 
\EQ
*j_{D3}^{Page} = *j_{D3}^{bs} - B\wedge *j_{D5}^{bs}
\label{tpage}
\EN
which means we define the RR Page current for D3-branes living
on the D5-brane in the way of compatible with T-duality.

\vs{5}
{\Large\bf 2.2 D3-brane in B fields}
\vs{2}

The Wess-Zumino term for D3-brane in the background of B fields is
\EQ
S_{WZ}^{D3} = \int \Big\{C_{4} + C_{2}\wedge (B+F) + \frac{1}{2}C_{0}
\wedge (B+F)\wedge (B+F) \Big\}
\label{d3wz}
\EN
which induces D3-, D1-brane and D-instanton sources, thus Eq.\p{bi3} is broken
and the nonvanishing
of modified Bianchi identity which describes the hodge dual of
the D3-brane source current is
\EQ
d*\tilde{F}_{5} + H\wedge\tilde{F}_{3} = d\tilde{F}_{5} + H\wedge\tilde{F}_{3}
= -*j^{bs}_{D3}
\label{nbi3}
\EN
where the brane source current can be derived by variation of
the  Wess-Zumino term for D3-brane \p{d3wz} with respect to $C_{4}$,
and we have taken the self-dual condition for $\tilde{F}_{5}$
into account. Eq.\p{d3wz} shows that the brane
source charges for D-strings and D-instantons in the D3-brane are
\bea
Q_{D1}^{bs} = \int *j_{D1}^{bs} &=& \int (B + F)\nn
Q_{D(-1)}^{bs} = \int *j_{D(-1)}^{bs} &=& \int\frac{1}{2}(B + F)\wedge(B + F)
\label{d3bs}
\ena
From the equations of motion \p{eqm}, the nonvanishing
Bianchi identity \p{nbi3} and two (modified) Bianchi identities
\p{bi5} and \p{bi7}, we see that $Q_{D3}^{bs}$
is conserved, but $Q_{D1}^{bs}$ and $Q_{D(-1)}^{bs}$ are not,
so we have to look for some conserved charges for D-strings and
D-instantons. Making use of the technique mentioned above Eq.\p{dieq},
from the equations of motion \p{eqm}, the nonvanishing
Bianchi identity \p{nbi3} and two (modified) Bianchi identities
\p{bi5} and \p{bi7}, we reach two equations whose
left sides are total derivative and right sides as localized
term
\EQ
d\Delta_{D1} = *j_{D1}^{bs} - B\wedge *j_{D3}^{bs}
\label{d31eq}
\EN
\EQ
d\Delta_{D(-1)} = *j_{D(-1)}^{bs} - B\wedge *j_{D1}^{bs}
+ \frac{1}{2}B\wedge B\wedge *j_{D3}^{bs}
\label{d3ieq}
\EN
where $\Delta_{D(-1)}$ is given in \p{deltai} and $\Delta_{D1}$
is defined as 
\EQ
\Delta_{D1} = -*F_{3} + B\wedge\tilde{F}_{5} + \frac{1}{2}B\wedge B\wedge
\tilde{F}_{3} + \frac{1}{6}B\wedge B\wedge B\wedge F_{1}
\label{delta1}
\EN
Eqs.\p{d31eq} and \p{d3ieq} show that
the hodge duals of Page currents for D-strings and D-instantons
can be consistently defined
\EQ
*j_{D1}^{Page} = *j_{D1}^{bs} - B\wedge *j_{D3}^{bs}
\label{d31c}
\EN
\EQ
*j_{D(-1)}^{Page} = *j_{D(-1)}^{bs} - B\wedge *j_{D1}^{bs} 
+ \frac{1}{2}B\wedge B\wedge *j_{D3}^{bs}
\label{d3ic}
\EN
From Eqs.\p{d3bs} and \p{d31c} and \p{d3ic}
the RR Page charges for D-strings and D-instantons
are
\bea
Q_{D1}^{Page} &=& \int *j_{D1}^{Page} = \int F\nn
Q_{D(-1)}^{Page} &=& \int *j_{D(-1)}^{Page}
= \int\Big[\frac{1}{2}(B+F)\wedge (B+F) - B\wedge (B+F) + 
\frac{1}{2}B\wedge B\Big]\nn
&=& \int\frac{1}{2}F\wedge F
\label{d31ip}
\ena
where we have seen the B-dependent terms 
$B\wedge B$, $B\wedge F$ have been cancelled with
each other. In~\cite{bds} and~\cite{dm}, it was argued that 
the RR Page charge for D1-branes in D3-brane defined 
by $\int\limits_{S_{2}}F$ should be quantized~\cite{kz}. 
In the context of WZW model, it has been shown that the 
integral $\int\limits_{S_{2}}F$ is quantized indeed~\cite{kkz}.

\vs{5}
{\Large\bf 2.3 D5-brane in B fields}
\vs{2}

We turn to a D5-brane in B fields,
the Wess-Zumino terms for D5-brane is 
\bea
S_{WZ}^{D5} &=& \int \Big\{C_{6} + C_{4}\wedge (B+F) + \frac{1}{2}C_{2}
\wedge (B+F)\wedge (B+F)\nn
&&
+ \frac{1}{6}C_{0}
\wedge (B+F)\wedge (B+F)\wedge (B+F)\Big\}
\label{d5wz}
\ena
which describes D5/D3/D1/D(-1) bound state. In the presence of D5-brane
source, the modified Bianchi identity \p{bi5} turn to be nonzero\footnote{In
IIB supergravity, the RR gauge field $C_{2p}$ is dual to $C_{(8-2p)}$.
Because of $**A_{2p+1} = -A_{2p}$ in 10D Euclidean spacetime~\cite{jp}, 
there are two 
ways to performing the electromagnetic duality. For example, consider
D5-brane RR gauge field $C_{6}$ (for illustration, we assume B fields 
vanish), we can define either ~~$*dC_{6} = dC_{2}$, i.e., 
$*dC_{2} = -dC_{6}$, or ~~$*dC_{2} = dC_{6}$, i.e., 
$*dC_{6} = -dC_{2}$. In case 1, the D5-brane source current is 
described by $dF_{3} = *j^{bs}_{D5}$, and in case 2, it is 
$dF_{3} = -*j^{bs}_{D5}$. The WZ term does not contain any
information about which definition one should use~\cite{dl}. In order
to make the definition for RR Page current compatible with T-duality,
we have to choose $dF_{3} = *j^{bs}_{D5}$.}
\EQ
d\tilde{F}_{3} + H\wedge F_{1} = *j^{bs}_{D5}
\label{nbi5}
\EN
Rewriting the term $H\wedge\tilde{F}_{3}$ in \p{nbi3} as
$d(B\wedge\tilde{F}_{3}) - B\wedge d\tilde{F}_{3}$ and inserting \p{nbi5}
into $B\wedge d\tilde{F}_{3}$, besides the exterior derivative
and localized terms, we have the term $\frac{1}{2}d(B\wedge B)\wedge F_{1}$
which can written as $\frac{1}{2}d(B\wedge B\wedge F_{1}) - 
\frac{1}{2}B\wedge B\wedge dF_{1}$. Exploiting the Bianchi identity
for $F_{1}$ \p{bi7}, we get the new
equation whose left side is exterior derivative and the right side
is localized term
\EQ
d\Delta_{D3} = *j^{bs}_{D3} - B\wedge *j^{bs}_{D5}
\label{d53eq}
\EN
with
\EQ
\Delta_{D3} = -*\tilde{F}_{5} - B\wedge\tilde{F}_{3} - 
\frac{1}{2}B\wedge B\wedge F_{1}
\label{delta3}
\EN
which shows that the properly defined RR Page current is 
\EQ
*j_{D3}^{Page} = *j_{D3}^{bs} - B\wedge *j_{D5}^{bs}
\label{d53c}
\EN
Here we should point out that Eq.\p{d53c} is derived from the
nonvanishing modified Bianchi identity \p{nbi3} for D3-brane, 
the other nonvanishing Bianchi identity \p{nbi5} for D5-brane,
and the Bianchi identity for $F_{1}$ \p{bi7}
\footnote{If we choose $d\tilde{F}_{3} + H\wedge F_{1} = -*j^{bs}_{D5}$
instead of Eq.\p{nbi5}, then we would have $*j_{D3}^{Page} = 
*j_{D3}^{bs} + B\wedge *j_{D5}^{bs}$,
which is incompatible with T-duality.}.
On the other hand, Eq.\p{tpage} is obtained from T-duality, which
means the ambiguity in the definition of $C_{6}$ can be fixed by
T-duality.
By exploiting of the equations of motion \p{eqm}, two nonvanishing modified
Bianchi identities \p{nbi3} and \p{nbi5}, 
the Bianchi identity \p{bi7}, and in the similar way used 
in subsection 2.1, we get the relations
\bea
d\Delta_{D1} &=& *j_{D1}^{bs} - B\wedge *j_{D3}^{bs} + 
\frac{1}{2}B\wedge B\wedge *j_{D5}^{bs}\nn
d\Delta_{D(-1)} &=& *j_{D(-1)}^{bs} - B\wedge *j_{D1}^{bs} + 
\frac{1}{2}B\wedge B\wedge *j_{D3}^{bs} - 
\frac{1}{6}B\wedge B\wedge B\wedge *j_{D5}^{bs}
\label{d51eq}
\ena
where $\Delta_{D1}$ and $\Delta_{D(-1)}$ are defined in \p{delta1} and
\p{deltai}. Eq.\p{d51eq} indicates that the hodge dual of the RR Page currents 
for D-strings and D-instantons should be defined as
\EQ
*j_{D1}^{Page} = *j_{D1}^{bs} - B\wedge *j_{D3}^{bs} + 
\frac{1}{2}B\wedge B\wedge *j_{D5}^{bs}
\label{d51c}
\EN
\bea
*j_{D(-1)}^{Page} &=& *j_{D(-1)}^{bs} - B\wedge *j_{D1}^{bs}
+ \frac{1}{2}B\wedge B\wedge
*j_{D3}^{bs}\nn
&&
 - \frac{1}{6}B\wedge B\wedge B\wedge *j_{D5}^{bs}
\label{d5ic}
\ena
Recall that for D5/D3/D1/D(-1) bound state, the brane sources
can be read off from the WZ term for D5-brane \p{d5wz}, the 
corresponding brane source charges are
\bea
Q_{D3}^{bs} &=& \int *j_{D3}^{bs} = \int (B+F)\nn
Q_{D1}^{bs} &=& \int *j_{D1}^{bs} = \int \frac{1}{2}(B+F)\wedge(B+F)\nn
Q_{D(-1)}^{bs} &=& \int *j_{D(-1)}^{bs} = \int \frac{1}{6}(B+F)\wedge(B+F)
\wedge(B+F)
\label{d5bs}
\ena
Inserting \p{d5bs} into \p{d53c}, \p{d51c} and \p{d5ic}, we obtain
the Page charges for D3-, D1-branes 
\EQ
Q_{D3}^{Page} = \int *j_{D3}^{Page} = \int F
\label{d53p}
\EN
\bea
Q_{D1}^{Page} &=& \int \Big(*j_{D1}^{bs} - B\wedge *j_{D3}^{bs} + 
\frac{1}{2}B\wedge B\wedge *j_{D5}^{bs}\Big)\nn
&=& \int \Big\{\frac{1}{2}(B+F)\wedge (B+F) - B\wedge (B+F) + \frac{1}{2}
B\wedge B\Big\}\nn
&=& \int \frac{1}{2}F\wedge F
\label{d51p}
\ena
and for D-instantons
\bea
Q_{D(-1)}^{Page} &=& \int \Big(*j_{D(-1)}^{bs} - B\wedge *j_{D1}^{bs}
+ \frac{1}{2}B\wedge B\wedge *j_{D3}^{bs}
- \frac{1}{6}B\wedge B\wedge B\wedge *j_{D5}^{bs}\Big)\nn
&=& \int \Big\{\frac{1}{6}(B+F)\wedge (B+F)\wedge (B+F)
- \frac{1}{2}B\wedge (B+F)\wedge (B+F)\nn
&& 
+ \frac{1}{2}B\wedge B\wedge (B+F) - \frac{1}{6}B\wedge B\wedge B\Big\}\nn
&=& \int\frac{1}{6}F\wedge F\wedge F
\label{d5ip}
\ena
Eq.\p{d5ip} shows that the terms $B\wedge B\wedge B$, $B\wedge B\wedge F$,
$B\wedge F\wedge F$ in the RR Page charge
of D-instantons living in D5-brane are cancelled with each other
nontrivially.

\vs{5}
{\Large\bf 2.4 D7-brane in B fields}
\vs{2}

The Wess-Zumino term for D7-brane in B fields is
\bea
S_{WZ}^{D7} &=& \int \Big\{C_{8} + C_{6}\wedge (B+F) + \frac{1}{2}C_{4}
\wedge (B+F)\wedge (B+F)\nn
&&
+ \frac{1}{6}C_{2}\wedge (B+F)\wedge (B+F)\wedge (B+F)\nn
&&
+ \frac{1}{24}C_{0}\wedge (B+F)\wedge (B+F)\wedge (B+F)\wedge (B+F)\Big\}
\label{d7wz}
\ena
which shows that the brane source charges for D5-, D3-,
D1-branes, and D-instantons in D7-brane, are
\bea
Q_{D5}^{bs} &=& \int *j_{D5}^{bs} = \int (B+F)\nn
Q_{D3}^{bs} &=& \int *j_{D3}^{bs} = \int \frac{1}{2}(B+F)\wedge(B+F)\nn
Q_{D1}^{bs} &=& \int *j_{D1}^{bs} = \int \frac{1}{6}(B+F)\wedge(B+F)
\wedge(B+F)\nn
Q_{D(-1)}^{bs} &=& \int *j_{D(-1)}^{bs} = \int \frac{1}{24}(B+F)\wedge(B+F)
\wedge(B+F)\wedge(B+F)
\label{d7bs}
\ena
In the presence of the D7-brane source, the Bianchi identity \p{bi7}
becomes nonvanishing one which describes the hodge dual of 
the D7-brane source current
\EQ
dF_{1} = -*j_{D7}^{bs}
\label{nbi7}
\EN
In the same principle as in the subsection 2.1, from
the equations of motion \p{eqm} and the nonvanishing (modified)
Bianchi identity \p{nbi3}, \p{nbi5} and \p{nbi7}, we have
\bea
d\Delta_{D5} &=& *j_{D5}^{bs} - B\wedge *j_{D7}^{bs}\nn
d\Delta_{D3} &=& *j_{D3}^{bs} - B\wedge *j_{D5}^{bs} + 
\frac{1}{2}B\wedge B\wedge *j_{D7}^{bs}\nn
d\Delta_{D1} &=& *j_{D1}^{bs} - B\wedge *j_{D3}^{bs} + 
\frac{1}{2}B\wedge B\wedge *j_{D5}^{bs}\nn 
&&
- \frac{1}{6}B\wedge B\wedge B\wedge *j_{D7}^{bs}\nn
d\Delta_{D(-1)} &=& *j_{D(-1)}^{bs} - B\wedge *j_{D1}^{bs} + 
\frac{1}{2}B\wedge B\wedge *j_{D3}^{bs}\nn
&&
- \frac{1}{6}B\wedge B\wedge B\wedge *j_{D5}^{bs}
+ \frac{1}{24}B\wedge B\wedge B\wedge B\wedge *j_{D7}^{bs}
\label{d7eq}
\ena
with
\EQ
\Delta_{D5} = \tilde{F}_{3} + B\wedge F_{1}
\EN
Eq.\p{d7eq} shows that the hodge duals of the RR Page currents for 
the D5-, D3-, D1-branes, and D-instantons in the D7-brane can be defined as
\bea
*j_{D5}^{Page} &=& *j_{D5}^{bs} - B\wedge *j_{D7}^{bs}\nn
*j_{D3}^{Page} &=& *j_{D3}^{bs} - B\wedge *j_{D5}^{bs}
+ \frac{1}{2}B\wedge B\wedge *j_{D7}^{bs}\nn
*j_{D1}^{Page} &=& *j_{D1}^{bs} - B\wedge *j_{D3}^{bs}
+ \frac{1}{2}B\wedge B\wedge *j_{D5}^{bs}\nn 
&&
- \frac{1}{6}B\wedge B\wedge B\wedge *j_{D7}^{bs}\nn
*j_{D(-1)}^{Page} &=& *j_{D(-1)}^{bs} - B\wedge *j_{D1}^{bs}
+ \frac{1}{2}B\wedge B\wedge *j_{D3}^{bs}\nn 
&&
- \frac{1}{6}B\wedge B\wedge B\wedge *j_{D5}^{bs}
+ \frac{1}{24}B\wedge B\wedge B\wedge B\wedge *j_{D5}^{bs}
\label{d7c}
\ena
Plugging Eq.\p{d7bs} into \p{d7c}, the Page charges are given by
\EQ
Q_{D5}^{Page} = \int *j_{D5}^{Page} = \int F
\EN
\bea
Q_{D3}^{Page} &=& \int\Big(*j_{D3}^{bs} - B\wedge *j_{D5}^{bs} + 
\frac{1}{2}B\wedge B\wedge *j_{D7}^{bs}\Big)\nn 
&&
\int \Big\{\frac{1}{2}(B+F)\wedge (B+F) - B\wedge (B+F) + \frac{1}{2}
B\wedge B\Big\}\nn 
&=& \int \frac{1}{2}F\wedge F
\ena
\bea
Q_{D1}^{Page} &=& \int \Big(*j_{D1}^{bs} - B\wedge *j_{D3}^{bs}
+ \frac{1}{2}B\wedge B\wedge *j_{D5}^{bs}
- \frac{1}{6}B\wedge B\wedge B\wedge *j_{D7}^{bs}\Big)\nn
&=& \int \Big\{\frac{1}{6}(B+F)\wedge (B+F)\wedge (B+F)
- \frac{1}{2}B\wedge (B+F)\wedge (B+F)\nn
&& 
+ \frac{1}{2}B\wedge B\wedge (B+F) - \frac{1}{6}B\wedge B\wedge B\Big\}\nn
&=& \int\frac{1}{6}F\wedge F\wedge F
\label{d71p }
\ena 
\bea
Q_{D(-1)}^{Page} &=& \int \Big(*j_{D(-1)}^{bs} - B\wedge *j_{D1}^{bs}
+ \frac{1}{2}B\wedge B\wedge *j_{D3}^{bs}
- \frac{1}{6}B\wedge B\wedge B\wedge *j_{D5}^{bs}\nn
&& 
+ \frac{1}{24}B\wedge B\wedge B\wedge B\wedge *j_{D7}^{bs}\Big)\nn
&=& \int \Big\{\frac{1}{24}(B+F)\wedge (B+F)\wedge (B+F)\wedge (B+F)\nn
&& 
- \frac{1}{6}B\wedge (B+F)\wedge (B+F)\wedge (B+F)
+ \frac{1}{4}B\wedge B\wedge (B+F)\wedge (B+F)\nn 
&&
- \frac{1}{6}B\wedge B\wedge B\wedge (B+F)
+ \frac{1}{24}B\wedge B\wedge B\wedge B\Big\}\nn
&=& \int\frac{1}{24}F\wedge F\wedge F\wedge F
\label{d7ip }
\ena
Eq.\p{d7ip } shows that all the B-dependent terms
like $B\wedge B\wedge B\wedge B $, $B\wedge B\wedge B \wedge F$,
$B\wedge B\wedge F\wedge F$, and $B\wedge F\wedge F\wedge F$
in the RR Page charge of D-instantons living in the D7-brane are 
cancelled with each other in intricate way.

\vs{5}
{\Large\bf 3. D3-brane probe in topologically nontrivial B fields}
\vs{2}

In section 2, we have constructed the conserved RR Page charges for 
D-branes in B fields, where we have exploited the properties of
Page charges -- conserved and localized. However, in the construction
we have used the assumption that in the whole region in which
D-branes move, the relation $H = dB$ holds. In this section,
we study how the RR Page charge of D-brane moving in 
topologically nontrivial B fields behaves. In the context of supergravity,
the simple case for brane creation effect~\cite{hw}-\cite{kz} is
D3-brane probe in the background of $\kappa$ coinciding NS5-branes
~\cite{op}-\cite{lrs} where there is topologically nontrivial B field.

The background fields around a stack of $\kappa$ coinciding 
flat NS5-branes are given by~\cite{chs}
\bea
ds^{2}&=&dx^{2} + f dy^{2},\nn
e^{2\Phi }&=&g_{s}^{2}f,\nn
H_{klm}&=&-\epsilon _{klmn}\partial _{n}f
\label{ns5}
\ena
where $\{ x^{\mu}\}=(x^{0}, x^{1}, \cdots \cdots x^{5})$ 
parameterize the directions along the NS5-branes, 
$\{y^{m}\}=(y^{6}, y^{7}, y^{8}, y^{9})$ are locations of 
the NS5-branes, and $g_{s}$ is the string coupling far from 
the branes. The harmonic function $f$ depends on the transverse 
space 
\bea
f&=&1+\frac{kl _{s}^{2}}{r^{2}}\nn
r&=&|\vec{y}| =\sqrt{k}l _{s}e^{\phi }. 
\label{ya}
\ena
In the spherical coordinates, the metric and $H$ turn into
\bea
ds^{2}&=&dx^{2} + fr^{2}(d\phi ^{2}+d\Omega _{3}^{2}),\nn
H&=&2kl _{s}^{2}\omega _{3}, 
\label{sph}
\ena
where $d\Omega _{3}^{2}$ and $\omega _{3}$ are the metric 
and volume form on the unit 3-sphere $S^{3}_{6789}$. The NS
3-form field strength $H$ satisfies $dH = 0$ and 
$\int\limits_{S^{3}}H\neq 0$.
For the following discussions, we choose 
the cylindrical coordinates $(z, \rho , \theta , \varphi )$
\bea
(y^{6}, y^{7}, y^{8}, y^{9})
=(z, \rho {\rm cos}\theta , \rho {\rm sin}\theta {\rm cos} \varphi,
\rho {\rm sin}\theta {\rm sin} \varphi)
\ena
and spherical coordinates to replace $(z, \rho )$ by
\bea
(z, \rho )=(r{\rm cos}\psi , r{\rm sin}\psi )
\ena
where $\theta \in [0, \pi]$, $\varphi \sim \varphi +2\pi $, $\psi 
\in [0, \pi]$. 

Consider a D3-brane probe in the background \p{ns5}, the 
D3-brane action is 
\EQ
S_{D3} = -T_{3}\int e^{-\Phi}\sqrt{-det(G_{ab} + B_{ab} + F_{ab})}
+ \int \Big(C_{4} + C_{2}\wedge (B+F)\Big)
\label{d3b}
\EN
Here we assume $(B+F)\wedge (B+F) = 0$, so there are only D-strings
living in the D3-brane, but no D-instantons. The equations of motion
of the D3-brane in the background \p{ns5} can be derived by the variation
of the action \p{d3b}. The BPS equation is~\cite{op}
\EQ
\frac{dz}{d\rho} = -\frac{1}{1 + e^{2\phi}}\frac{\psi - \psi_{0} -
\frac{1}{2}\sin2\psi}{\sin^{2}\psi}
\label{bps}
\EN
The solution of the BPS equation \p{bps} gives the shape of the D3-brane.
The typical feature for this BPS D3-brane configuration is that it
includes an infinite tube which can be interpreted as D1-brane.
The angle $\psi_{0}$ has a simple geometrical meaning:
opening angle as shown in figure 1. Especially when $z_{max}\rightarrow 
\infty$, the infinite tube can be identified with D-strings~\cite{op} and
\cite{lrs}.

\begin{figure}[htbp]
\begin{center}
\includegraphics*[scale=.60]{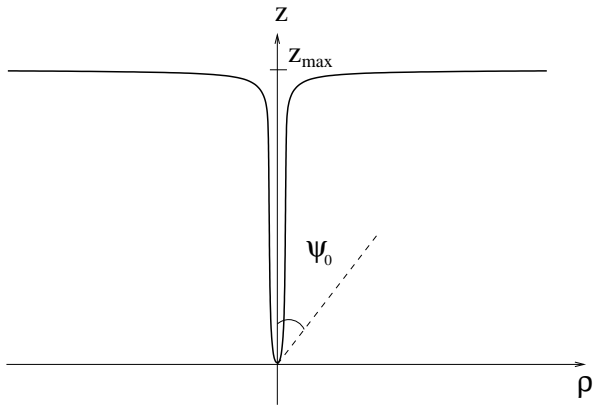}
\vspace{-0.7cm}
\end{center}
\caption{D3-brane profile for the fixed large 
$z_{{\rm max}}$ with opening angle $0< \psi_{0} < \pi /2$}
\end{figure}


The 2-form $B+F$ on the D3-brane is  given~\cite{bds},\cite{op},\cite{lrs}
\EQ
{\cal F} = B+F = \kappa l_{s}^{2}(\psi - \frac{1}{2}\sin2\psi - \psi_{0})
\sin\theta d\theta\wedge d\varphi
\label{bf}
\EN
from the discussion in section 2, we know $\int\limits_{S^{2}}{\cal F}$
is D1-brane source charge and not conserved.

When the observer stays at the same side of D3-brane, i.e., $z > 0$,
one can choose the gauge in which the NS B-field is proportional to
the volume form of the two-sphere spanned by $(\theta, \varphi)$~\cite{bds}
and \cite{op}
\EQ
B = \kappa l_{s}^{2}(\psi - \frac{1}{2}\sin2\psi)
\sin\theta d\theta\wedge d\varphi
\label{bs}
\EN 
which is the smooth choice everywhere except at the point 
$\psi = \pi$ ($z < 0$). In general, if the observer is at one side of
the NS5-branes, the proper gauge choice for NS B-field is that
its singular point should be at the other side of NS5-branes.
Eq.\p{bs} shows that the singular point $\psi = \pi$ is indeed
at the opposite side of the observer relative to NS5-branes $z < 0$.
From Eq.\p{bf} and \p{bs}, the corresponding $U(1)$ gauge field strength
$F$ is
\EQ
F = -\kappa l_{s}^{2}\psi_{0}
\label{fs}
\EN
Here we should mention that the parameter $\psi_{0}$ should be 
quantized not only
from WZW model~\cite{kkz} but also from supergravity consideration~\cite{cpr}
\EQ
\psi_{0} = n\frac{\pi}{\kappa}
\label{quan}
\EN
The RR Page charge of D1-branes on the D3-brane in the gauge \p{bs} is 
\EQ
Q^{Page}_{D1} = \frac{1}{(2\pi l_{s})^{2} }\int\limits_{S^{2}}F 
= -\frac{\kappa}{(2\pi)^{2}}\int\limits_{S^{2}}\psi_{0} = -n
\label{qs}
\EN
For definitness, we assign the positive charge of D1-branes to represent
that D1-branes emanate from the D3-brane to the NS5-branes, then
Eq.\p{qs} can be described by figure 2, which means if the observer
stays at the same side of the D3-brane ($z > 0$) relative to the NS5-branes, 
he/she measures $n$ D1-branes
directing away from the NS5-branes and ending to the flat D3-brane.

\begin{figure}[htbp]
\begin{center}
\includegraphics*[scale=.60]{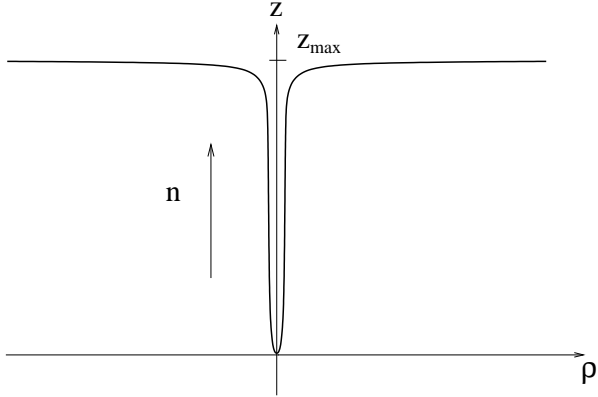}
\vspace{-0.7cm}
\end{center}
\caption{When the observer is at the same
side of the D3-brane, he/she
measures $n$ D1-branes emanate from
the NS5-branes to the D3-brane.}
\end{figure}

If the observer is at the opposite side of the D3-brane, i.e., $z < 0$,
one has to choose the other gauge
\EQ
B = \kappa l_{s}^{2}(\psi - \pi - \frac{1}{2}\sin2\psi)
\sin\theta d\theta\wedge d\varphi
\label{bn}
\EN
which is singular at $\psi = 0$ but smooth at $\psi = \pi$. 
Because $B+F$ is gauge invariant, two different
choice of the NS B-field result in different $F$.
From Eqs.\p{bf} and \p{bn}, the other $F$ is given
\EQ
F = \kappa l_{s}^{2}(\pi - \psi_{0})
\label{fn}
\EN
Then the RR Page charge of D1-branes in the other gauge is
\EQ
Q^{Page}_{D1} = \frac{1}{(2\pi l_{s})^{2}}\int\limits_{S^{2}}F 
= -\frac{\kappa}{(2\pi)^{2}}\int\limits_{S^{2}}(\pi - \psi_{0}) = \kappa - n
\label{qn}
\EN
which indicates that if the observer stays at the opposite side of
the D3-brane ($z < 0$), he/she measures $\kappa - n$ D1-branes direct away
from NS5-branes and end to the flat D3-brane which is illustrated
in figure 3.

\begin{figure}[htbp]
\begin{center}
\includegraphics*[scale=.60]{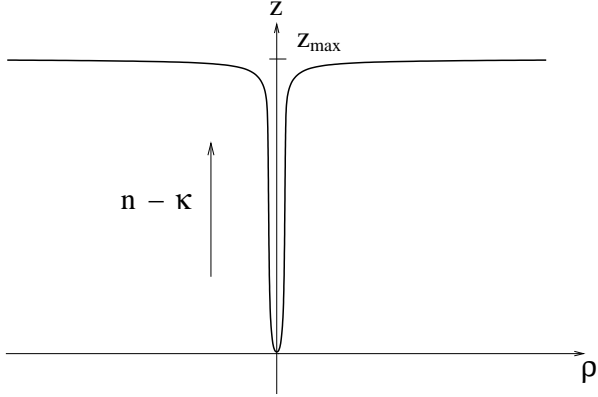}
\vspace{-0.7cm}
\end{center}
\caption{ When the observer is at the opposite
side of the D3-brane ($z < 0$), he/she
measures $\kappa - n$ D1-branes emanate from
the NS5-branes to the D3-brane.}
\end{figure}

Under the reversal $z \rightarrow -z$, that is, if
we reassign the observer stays at the side of $z > 0$
but the asymptotically flat D3-brane is at $z_{max} \rightarrow -\infty$,
we have figure 4 which is equivalent to figure 3.
In the context of WZW model, Fig.2 corresponds to the Cardy
boundary state $|n>_{c}$, the reversal $z \rightarrow -z$
is realized by the rotation operator 
$exp\{i\pi (J^{3}_{0} - \bar{J}^{3}_{0})\}$.
Acting the rotation operator on the Cardy
boundary state $|n>_{c}$, we have $exp\{i\pi (J^{3}_{0} - \bar{J}^{3}_{0})\}
|n>_{c} = |\kappa - n>_{c}$ which is described by Fig.4~\cite{kz}.

\begin{figure}[htbp]
\begin{center}
\includegraphics*[scale=.60]{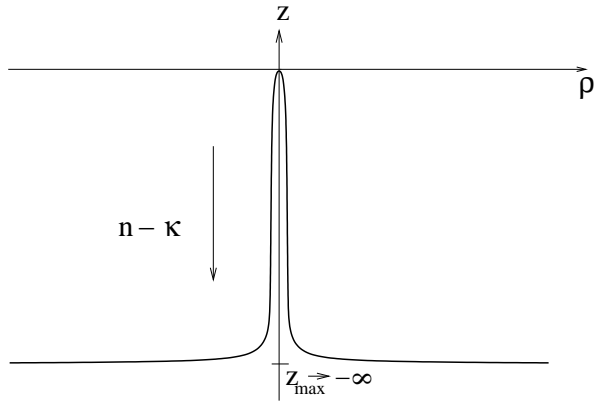}
\vspace{-0.7cm}
\end{center}
\caption{The other equivalent description of 
figure 3, if we assume the observer
stays at the side of $z > 0$
but the flat D3-brane is at 
$z_{max} \rightarrow -\infty$.}
\end{figure}

Combining figure 4 and figure 2, the observer (staying
in the region $z > 0$) has the following physical
picture: when $t \rightarrow -\infty$, the D3-brane initially locates at  
$z_{max} \rightarrow -\infty$ where there are 
$n - \kappa$ D1-branes emanating from the NS5-branes to the 
D3-brane, late the lower D3-brane passes through the 
$\kappa$ coinciding NS5-branes and finally stays at 
$z_{max} \rightarrow \infty$ but there are 
$n$ D1-branes directing away the NS5-branes and ending
on the D3-brane, so in the whole physical process, 
$\kappa$ D1-branes have been created which is nothing
but the Hanany-Witten effect~\cite{hw}-\cite{kz}. In the
above discussion, we find that in the topologically nontrivial
B-field, the RR Page charge is no longer to be conserved,
instead there is a jump in the RR Page charge of the D1-branes
when the lower D3-brane crosses the NS5-branes, and it is
this jump which describes the brane creation phenomenon.

In the above discussion, we see that due to the
topological nontriviality of the NS B-field, there are two different
gauge choices for it, one is singular at the south pole $\psi = \pi$
and the other is singular at the north pole $\psi = 0$.
Each gauge choice for NS B-field induces different $U(1)$ gauge
field strength F which is related to the number of the
D1-branes suspending between the NS5-branes and asympototically
flat D3-brane. The jump between two different F can be
interpreted as brane creation effect. It is reminiscent of
the string creation in the system of the D2-brane probe and the 
D6-brane background which can be obtained by dimensional
reduction from 11D Taub-NUT solution~\cite{my}, its potential is
\EQ
A^{S}_{\varphi} = \frac{R}{2}(1 + \cos\theta)
\EN
where R is the radius of the circle of the eleventh dimension,
and the superscript S denotes the coordinates regular at the
south pole $\theta = \pi$. The metric of Taub-NUT solution
can be chosen to be nonsingular at the south pole but
singular at the north pole $\theta = 0$. The singularity
at the north pole can be shifted to the south pole by
the gauge transformation
\EQ
x^{10}_{N} = x^{10}_{S} + R\varphi
\label{gt}
\EN
thus the direction of Dirac string is changed and the
potential turns into
\EQ
A^{N}_{\varphi} = \frac{R}{2}(-1 + \cos\theta)
\EN
By proper embedding of an M2-brane in the 11D Taub-NUT
configuration, it was found that the M2-brane has no
winding around eleventh-direction in the coordinate
system regular at the south pole, but it can wind
around this direction indeed in the coordinate system
regular at the north pole obtained by the gauge
transformation \p{gt}, which implies string charge
creation from the ten-dimensional point of view~\cite{my}.

In section 2, we have seen that when the NS B-field is not constant,
i.e., $H = dB \neq 0$, the RR brane source charge is not conserved.
Now we survey whether the nonconservation
of the brane source charge encodes any information about
Hanany-Witten effect or not. The above D3-brane probe in
the NS5-brane background is an ideal laboratory to implement
this idea. The equations of motion and (modified) Bianchi identities 
which describe the D3-brane in the background of the NS5-branes
are given by 
\bea
d*F_{1} - H\wedge *\tilde{F}_{3} &=& 0\nn
d*\tilde{F}_{3}- H\wedge\tilde{F}_{5} &=&-*j^{bs}_{D1}\nn
d*\tilde{F}_{5} + H\wedge \tilde{F}_{3} &=& 
d\tilde{F}_{5} + H\wedge \tilde{F}_{3} = -*j^{bs}_{D3}\nn
d\tilde{F}_{3} + H\wedge F_{1} &=& 0\nn
dF_{1} &=& 0
\label{teq}
\ena
In the spherical coordinates, the NS H field strength for the 
$\kappa$ coincident NS5-branes satisfies 
\bea
dH = 0, ~~~ \int\limits_{S^{3}}H\neq 0
\label{hp}
\ena
By exploiting of Eqs.\p{teq} and \p{hp}, we find that
\EQ
d*j^{bs}_{D3} = 0
\EN
\EQ
d*j^{bs}_{D1} = H\wedge *j^{bs}_{D3}
\label{nc}
\EN
which indicates that the brane source
charge for the D3-brane is conserved, but for D1-branes it is not.
In order to see what the physical meaning of Eq.\p{nc} is and whether
it encodes the Hanany-Witten effect, we integrate \p{nc} over 
nine dimensional manifold $M_{9}$ and have
\bea
\int\limits_{M_{9}}d*j^{bs}_{D1} = \int\limits_{M_{9}\bigcap D3}H
= \int\limits_{S^{3}}H
\label{nci}
\ena
Since $\Big(j^{bs}_{D1}(x)\Big)^{\mu\nu} 
\sim\int d^{4}\xi\epsilon^{abcd}\pa_{a}
X^{\mu}\pa_{b}X^{\nu}\Big(B_{\rho\sigma}\pa_{c}X^{\rho}\pa_{d}X^{\sigma} 
+ F_{cd}\Big)\delta^{10}(x - X)$, it is easy to check 
$d*j^{bs}_{D1}\sim d(B+F) = H$, and Eq.\p{nci} is a sort of identity
$d(B+F) = H$. Then we find that the nonconservation of the 
brane source charge of the D1-branes \p{nc} reflects the fact that
$dB = H\neq 0$, which has nothing to do with Hanany-Witten effect.

\vs{5}
{\Large\bf 4. Summary and Conclusion}
\vs{2}

In the above, we have derived the
RR Page charges for D(2p+1)-branes with topologically trivial
B-field from brane source charges in type IIB supergravity.
We have considered
IIB supergravity plus D-brane sources, from which we have obtained 
the equations
of motion and the nonvanishing (modified) Bianchi identities that
define the duals of brane source currents for D(2p+1)-branes.
By inserting the equations for the
duals of brane source currents for D(2p+1+2n)-branes ($n > 1$) into
that for the D(2p+1)-brane iteratively, we have found that the resulting
equations can be recast into the form whose left sides of equations
are exterior derivative and the right sides are localized objects,
which indicates that the right side localized objects can be identified
as Page charges because of their conservation and locality. 
Since there are two types of Chern-Simons terms in supergravity 
theories, one is the wedge product of one potential with two
field strengths which induces the equation of motion for gauge fields
with the universal form $df_{8-2p}\pm H\wedge f_{5-2p} = -*j_{2p+2}$
\footnote{For D7-brane source, it is $dF_{2}= -*j_{D7}$,
but in our case it is the last equation for the dual of
brane source current.},
the other appears in the kinetic term for the modified field
strengths which results in nonvanishing of modified Bianchi identities
with the same forms as the equations of motion. 
The universal forms $df_{8-2p}\pm H\wedge f_{5-2p} = -*j_{2p+2}$
(for the equations of motion and nonvanishing of modified Bianchi identities)
make it feasible to rewrite the equations in the form whose
left sides are exterior derivative and right side are localized
terms, which shows that the Page charges 
for D(2p+1)-branes can be consistently defined from the 
brane source charges indeed. 
Plugging 
the brane source charges into the expression of the Page charges,
we have shown that all the Page charges are independent of the background
B fields. In our explicit construction, it is highly nontrivial
that the B-dependent terms like $B\wedge B\wedge B\wedge B$,
$B\wedge B\wedge B\wedge F$, $B\wedge B\wedge F\wedge F$ 
and $B\wedge F\wedge F\wedge F$ from different sources
are exactly cancelled with each other. After discussing the
RR Page charges in topologically trivial B-field, 
we have turned to the topologically nontrivial case. In order to
study how Page charges behave in topologically
nontrivial B-field, we have considered D3-brane probe in the
background of $\kappa$ coincident NS5-branes.
With this example, we have shown that the nonconservation of brane
source charge does not describe to the Hanany-Witten effect, it
only reflects the fact that the brane source charge depends on
the background nonconstant B-field.
In the topological nontrivial B-field, we have found that the Page charge 
is no longer conserved, instead there is a jump between two Page 
charges defined in each patch, and we have interpreted 
this jump as Hanany-Witten effect.

\section*{Acknowledgement} 

I would like to thank D. Grumiller and M. Kreuzer
for valuable discussion. This work is supported
by the Austrian Research Funds FWF under grant No. M597-TPH.

\newcommand{\NP}[1]{Nucl.\ Phys.\ {\bf #1}}
\newcommand{\AP}[1]{Ann.\ Phys.\ {\bf #1}}
\newcommand{\PL}[1]{Phys.\ Lett.\ {\bf #1}}
\newcommand{\CQG}[1]{Class. Quant. Gravity {\bf #1}}
\newcommand{\CMP}[1]{Comm.\ Math.\ Phys.\ {\bf #1}}
\newcommand{\PR}[1]{Phys.\ Rev.\ {\bf #1}}
\newcommand{\PRL}[1]{Phys.\ Rev.\ Lett.\ {\bf #1}}
\newcommand{\PRE}[1]{Phys.\ Rep.\ {\bf #1}}
\newcommand{\PTP}[1]{Prog.\ Theor.\ Phys.\ {\bf #1}}
\newcommand{\PTPS}[1]{Prog.\ Theor.\ Phys.\ Suppl.\ {\bf #1}}
\newcommand{\MPL}[1]{Mod.\ Phys.\ Lett.\ {\bf #1}}
\newcommand{\IJMP}[1]{Int.\ Jour.\ Mod.\ Phys.\ {\bf #1}}
\newcommand{\JHEP}[1]{JHEP\ {\bf #1}}
\newcommand{\JP}[1]{Jour.\ Phys.\ {\bf #1}}


\begin{thebibliography}{99}
\bibitem{jp}
J. Polchinski, {\it String theory } (Cambridge University Press, 1998).

\bibitem{bds}
C. Bachas, M. Douglas and C. Schweigert, 
``{\it Flux stabilization of D-branes}", \JHEP{05} (2000) 048, 
hep-th/0003037.

\bibitem{wt}
W. Taylor,  ``{\it D2-branes in $B$ fields}", \JHEP{07} (2000) 039,
hep-th/0004141.

\bibitem{amm}
A. Alekseev, A. Mironov and A. Morozov, ``{\it On $B$-independence 
of RR charges} ", hep-th/0005244.

\bibitem{dm}
D. Marolf, ``{\it Chern-Simons terms and three notions of 
charge} ", hep-th/0006117.

\bibitem{jgz}
J.-G. Zhou, ``{\it D-branes in B Fields }", hep-th/0102178, 
to appear in Nucl. Phys. {\bf B}.

\bibitem{hw}
A. Hanany and E. Witten, 
``{\it Type IIB superstrings, BPS monopoles, and three dimensional 
gauge dynamics} ", Nucl. Phys. {\bf B 492} (1997) 152, hep-th/9611230.

\bibitem{bachas}
C. P. Bachas, M. R. Douglas and M. B. Green, 
``{\it Anomalous creation of branes} ", 
\JHEP{07} (1997) 002, hep-th/9705074.  

\bibitem{danielsson}
U. Danielsson, G. Ferretti and I.R. Klebanov, ``{\it Creation of 
fundamental strings by crossing D-branes} ", Phys. Rev. Lett. 
{\bf 79} (1997) 1984, hep-th/9705084.

\bibitem{bergman}
O. Bergman, M.R. Gaberdiel and G. Lifschytz, ``{\it Branes, 
orientifolds and the creation of elementary strings} ", Nucl. Phys. 
{\bf B 509} (1998) 194, hep-th/9705130.

\bibitem{dealwis}
S.P. de Alwis, ``{\it A note on brane creation} ", Phys. Lett. 
{\bf B 388} (1997) 720, hep-th/9706142.

\bibitem{osz}
N. Ohta, T. Shimizu and J.-G. Zhou, ``{\it Creation of fundamental 
string in M(atrix) theory} ", Phys. Rev. {\bf D 57} (1998) 2040, 
hep-th/9710218;
T. Kitao, N. Ohta and J.-G. Zhou, ``{\it Fermionic zero mode and 
string creation between D4-branes at angles} ", Phys. Lett. {\bf B 428} 
(1998) 68, hep-th/9801135.

\bibitem{callan1}
C.G. Callan, A. Guijosa and K.G. Savvidy, ``{\it Baryons and string 
creation from the fivebrane worldvolume action} ", Nucl. Phys. 
{\bf B 547} (1997) 127, hep-th/9810092.

\bibitem{my}
T. Matsuo and T.Yokono, ``{\it String creation in D6-brane
background}", \MPL {A 14} (1999) 1175, hep-th/9903081.

\bibitem{kz}
T. Kubota and J.-G. Zhou, ``{\it RR charges of D2-branes in group
manifold and Hanany-Witten effect}", \JHEP {12} (2000) 030, hep-th/0010170.

\bibitem{kkz}
A. Kling, M. Kreuzer and J.-G. Zhou, ``{\it $SU(2)$ WZW D-branes
and quantized worldvolume $U(1)$ flux on $S^{2}$ }", \MPL {A 15} 
(2000) 2069, hep-th/0005148.

\bibitem{dl}
M. Li, ``{\it Boundary states of D-branes and Dy-strings}", \NP {B460}
(1996) 351, hep-th/9510161; M. Douglas, ``{\it Branes within Branes}",
in Cargese 97: String, Branes and Dualities, p.267, hep-th/9512077.

\bibitem{op}
O. Pelc, ``{\it On the quantization constraints for a D3 brane 
in the geometry of NS5 branes }", \JHEP {08} (2000) 030, hep-th/0007100.

\bibitem{lrs}
P.M. Llatas, A.V. Ramallo and J.M. Sanchez de Santos,
``{\it World-volume solitons of the D3-brane in the background
of $(p,q)$ five-branes} ", hep-th/9912177.

\bibitem{chs}  
C.G. Callan Jr., J.A. Harvey and A. Strominger, ``{\it Supersymmetric 
string solitons} ", Lectures given by C. Callan and J. Harvey at the 
1991 Trieste spring school ``String theory and quantum gravity", 
hep-th/9112030.

\bibitem{cpr}
J.M. Camino, A. Paredes and A.V. Ramallo,
``{\it Stable wrapped branes} ", \JHEP {05} (2001) 011, hep-th/0104082.



\end{thebibliography}
\end{document}